\documentclass[a4paper,british]{jpconf}
\usepackage[T1]{fontenc}
\usepackage[latin9]{inputenc}
\usepackage{babel}

\usepackage{amstext}
\usepackage{amssymb}
\usepackage{esint}
\usepackage[unicode=true, 
 bookmarks=true,bookmarksnumbered=false,bookmarksopen=false,
 breaklinks=false,pdfborder={0 0 1},backref=false,colorlinks=false]
 {hyperref}
\hypersetup{pdftitle={The cosmological constant and the relaxed universe},
 pdfauthor={Florian Bauer},
 pdfsubject={PASCOS 2010 Talk: Relaxing a large cosmological constant},
 pdfkeywords={Cosmological constant, dark energy, modified gravity, relaxation, Palatini}}

\makeatletter

\usepackage{iopams}
\usepackage{setstack}

\makeatother

\begin{document}

\title{The cosmological constant and the relaxed universe}

\author{Florian Bauer}

\address{\textup{High Energy Physics Group, Dept.\ ECM, and Institut de Ci\`encies
del Cosmos,}\\
\textup{Universitat de Barcelona, Av.\ Diagonal 647, E-08028 Barcelona,
Catalonia, Spain}}

\ead{fbauerphysik@eml.cc}
\begin{abstract}
We study the role of the cosmological constant~(CC) as a component
of dark energy~(DE). It is argued that the cosmological term is in
general unavoidable and it should not be ignored even when dynamical
DE sources are considered. From the theoretical point of view quantum
zero-point energy and phase transitions suggest a CC of large magnitude
in contrast to its tiny observed value. Simply relieving this disaccord
with a counterterm requires extreme fine-tuning which is referred
to as the old CC problem. To avoid it, we discuss some recent approaches
for neutralising a large CC dynamically without adding a fine-tuned
counterterm. This can be realised by an effective DE component which
relaxes the cosmic expansion by counteracting the effect of the large
CC. Alternatively, a CC filter is constructed by modifying gravity
to make it insensitive to vacuum energy. 
\end{abstract}
\label{hiereinfachnurunsinn}%

\section{Introduction}

The old CC problem is of theoretical nature~\cite{Weinberg:1988cp}.
It is not forbidden to explain the current accelerated expansion of
the universe by a small CC or vacuum energy density~$\Lambda_{\text{obs}}\sim10^{-47}\,\text{GeV}^{4}$.
The cosmological term can be considered as the zero-order term in
almost every action functional and it is not absent unless further
assumptions are made. If it were a free parameter its observationally
favoured smallness would not pose any problem. And due to its minimalist
nature it would be a perfect candidate to explain the late-time expansion
history.

However, modern theories suggest the existence of large contributions
to the CC~$\Lambda$ which have to sum up to the tiny value~$\Lambda_{\text{obs}}$.
This is a highly non-trivial requirement because all those parts are
usually not related to each other and they can be of different magnitude
and sign. Perhaps the worst part comes from quantum field theory in
the form of vacuum or zero-point energy. Since every field mode provides
a contribution related to its energy~$\omega$, the integral over
all modes in the energy spectrum results into an infinite CC contribution.
For a massless field $\omega$ is just the mode momentum $p$ and
one finds the quartically divergent result \begin{equation}
\Lambda_{0}\sim\int_{0}^{\infty}d^{3}p\, p.\label{eq:Lam-ZeroPoint}\end{equation}
Common estimations impose an ultra-violet~(UV) cutoff at a high-energy
scale~$p_{\text{max}}$, which replaces the upper bound of the integral.
This naive procedure yields the often presented number $|\Lambda_{0}/\Lambda_{\text{obs}}|\sim10^{123}$
when the maximal momentum is assumed to be around the Planck mass~$M_{P}\sim10^{19}\,\text{GeV}$.
So far no evidence has been found for the existence of a cutoff indicating
that~$p_{\text{max}}$ should be at least above $M_{\text{ew}}\sim10^{2}\,\text{GeV}$,
which is roughly the energy scale accessible to current accelerator
experiments. Obviously, this value is still too high. The energy cutoff
compatible with~$\Lambda_{\text{obs}}$ would be of the order of
neutrino masses~$p_{\text{max}}\sim10^{-3}\,\text{eV}$, which is
in clear contrast to quantum field theory tested up to the electro-weak
scale~$M_{\text{ew}}$. However, it should be noted that vacuum energy%
\footnote{We mean its absolute value, not finite differences related to the
Casimir effect.%
} cannot be measured in the lab because it couples only to gravity.
Therefore, a low UV cutoff just for the zero-point energy part could
be allowed by observations, but it would be awkward from the theoretical
point of view. Finally, using renormalisation the divergence in~Eq.~(\ref{eq:Lam-ZeroPoint})
can be formally removed, but finite terms remain typically of the
order~$m^{4}$ for fields with mass~$m$. Despite the fact that
we do not know how to handle the term in~(\ref{eq:Lam-ZeroPoint}),
there is no reason to believe that it is small or zero. Instead, it
might be a large quantity according to the above considerations. 

Next, a more quantitative CC contribution comes from the symmetry
breaking of the electro-weak sector of the standard model. As the
Higgs field acquires a non-zero vacuum expectation value in the broken
phase the value of the potential in the new minimum is different from
its minimal value in the unbroken phase. Thus, the phase transition
induces a finite vacuum energy shift~$\mathcal{O}(M_{\text{ew}}^{4})$
again much larger than~$\Lambda_{\text{obs}}$. Apart from this classical
term, quantum corrections render the situation worse as new terms
occur at every order in perturbation theory~\cite{Bauer:2010wj}.

Another source of vacuum energy could be primordial inflation, where
a large positive value of the inflaton potential drives the rapid
accelerated expansion in the early universe. At the end of the inflationary
period the inflaton scalar approaches the minimum of its potential,
where the effective vacuum energy is much lower than before. However,
this does not imply that it is zero or of the observed CC value, it
could be much larger in magnitude and even negative. In that case
the well known Big Bang evolution had stopped already far in the past. 

It is unlikely that the sum~$\Lambda$ of all contributions mentioned
above is of the order~$\Lambda_{\text{obs}}$, instead it is expected
to be of the order of the largest contribution. To obtain consistency
between the theoretical value~$\Lambda$ and the observed one, the
easiest way is adding a counterterm~$\Lambda_{\text{ct}}$ which
matches~$(-\Lambda)$ to very high precision such that~$\Lambda_{\text{ct}}+\Lambda\approx\Lambda_{\text{obs}}$.
This procedure is considered to be highly unnatural since it requires
an enormous amount of fine-tuning in~$\Lambda_{\text{ct}}$, e.g.\
fixing up to~$122$ decimal digits. 

Motivated by the problems of understanding the zero-point energy in
Eq.~(\ref{eq:Lam-ZeroPoint}), it has become popular to {}``replace''
the CC by dynamical DE. Note that in general this does not address
the old CC problem. Instead it introduces more steps. First, any CC
originating e.g.\ from the sources mentioned above has to be removed
somehow. Clearly, without further explanation it means that a fine-tuned
counterterm is added which cancels~$\Lambda$. This step is often
suppressed in the description of DE models. Second, a new dynamical
source is introduced, which drives the late-time accelerated expansion~\cite{Copeland:2006wr}.
Obviously, the second step requires the first one but not vice-verse.
The reason is that making the CC vanish or much smaller than~$\Lambda_{\text{obs}}$
requires even more fine-tuning than adjusting it to match exactly
the observed value. On the other hand, one could argue that a yet
unknown theory of quantum gravity (or an undiscovered feature of the
known candidates) makes the CC vanish and clears the path for dynamical
DE models. However, this unknown but probably powerful theory could
also arrange the CC to match its observed value and no additional
dynamical DE component would be needed. Moreover, it seems that at
the moment observations do not require a dynamical nature of DE~\cite{Amanullah:2010vv}.
Of course, the situation might change in the future when observational
data becomes better and more precise. But even in that case the problem
of getting rid of a large $\Lambda$ persists and the CC as a static
DE component is by no way ruled out. Accordingly, the relevant question
about DE and the accelerated expansion is not whether there is either
a static CC or a dynamical component, but if we need only a CC or
a CC plus something dynamical. More complicated but completely analogous
to dynamical DE components are modified gravity theories with built-in
late-time acceleration~\cite{Nojiri:2006ri} or unconventional cosmological
models, where the observed acceleration is just an apparent effect.
It will be interesting to know whether nature exhibits more complexity
than the simplest explanation. In the {}``worst'' case we have to
deal with a mixture of many sources for acceleration in addition to
the old CC problem. 

In the following we concentrate on the latter problem, i.e.\ we accept
the existence of a presumed large value~$|\Lambda|\gg\Lambda_{\text{obs}}$
as a part of the energy content of the universe. Conventional energy
sources like dust and radiation dilute with the expansion of the universe,
and without counter-measures the large CC term eventually starts dominating
the cosmic expansion very early in the cosmic history. As a result,
$\Lambda<0$ initiates a collapse and the universe dies in a Big Crunch.
For~$\Lambda>0$ an inflationary epoch with a large expansion rate~$H$
starts and continues forever because the constant~$\Lambda$ does
not offer a graceful exit. In order to avoid these catastrophic scenarios
and to permit a reasonable Big Bang-style universe, we have to neutralise
the effect of~$\Lambda$. Of course, we want to avoid the introduction
of a fine-tuned counterterm, instead we look for a dynamical mechanism
or a suitable modification of gravity, which tames the large CC term.
Here, we discuss three recent approaches in this context. A relaxation
model with a variable CC (including the large term~$\Lambda$), a
similar model implemented as modified gravity in the metric formalism,
and finally a CC filter in a setup using the Palatini formalism. All
these models have in common that they do not need a fine-tuned counterterm
despite the existence of a large CC.

\section{Relaxing a large CC}

\label{sec:LXCDM}Let us begin with a simple example of the relaxation
mechanism. To neutralise the effect of the large CC we introduce an
additional component of opposite sign, which adjusts to the value
of~$\Lambda$ dynamically. In our first model the total DE density
is given by $\rho_{\Lambda}=\Lambda+\beta/f$ with a constant parameter~$\beta$
and a function~$f$. According to the Friedmann equation the Hubble
expansion rate~$H$ is sensitive to the sum of all energy densities
\begin{equation}
\rho_{c}=\frac{3H^{2}}{8\pi G_{N}}=\rho_{m}+\rho_{\Lambda}=\rho_{m}+\Lambda+\frac{\beta}{f},\label{eq:LXCDM-Friedmann}\end{equation}
where~$\rho_{c}$ is the critical energy density, $G_{N}$ is Newton's
constant, and~$\rho_{m}$ denotes the energy density of matter and
radiation. Without the $\beta/f$ term the constant~$\Lambda$ would
control the expansion when~$\rho_{m}$ becomes sufficiently small.
However, we want to have a well-behaved late-time evolution with~$H$
of the order of the currently observed Hubble rate~$H_{0}$. For
this purpose consider the dynamical term defined by~$f=H^{2}$ and~$\beta$
having the opposite sign of~$\Lambda$~\cite{Stefancic:2008zz}.
As a result of the small Hubble rate at late times only the term~$(\Lambda+\beta/H^{2})$
is relevant, which is obvious after dividing Eq.~(\ref{eq:LXCDM-Friedmann})
by $|\Lambda|\gg\rho_{c},\rho_{m}$,\begin{equation}
1+\frac{\beta}{\Lambda H^{2}}=\frac{\rho_{c}-\rho_{m}}{\Lambda}.\label{eq:LXCDM-rhoL-null}\end{equation}
The right-hand side is much smaller than unity and can be safely neglected.
Therefore, we find the consistent solution $H^{2}\rightarrow H_{e}^{2}=-\beta/\Lambda$
describing a de~Sitter cosmos with a low Hubble rate~$H_{e}$, which
complies with current observations. Without the term $\beta/f$ the
expected value of~$H^{2}$ would be proportional to $\Lambda$ and
thus very large. Here, $H_{e}^{2}$ is inversely proportional to the
large CC indicating a low Hubble rate. Note also the absence of fine-tuning
in the parameter~$\beta$. Small changes in~$\beta$ only lead to
small changes in~$H_{e}^{2}$, hence, it is not necessary to fix
a lot of decimal digits. 

The example~$f=H^{2}$ works well at late times when $H$ is small
such that the dynamical term~$\beta/H^{2}$ becomes large and dynamically
relaxes the large CC term. At earlier times in the cosmic history,
the Hubble rate was much larger and we have to enhance the function~$f$
for the relaxation mechanism to work. First, consider the radiation
era, where the scale factor~$a\propto t^{1/2}$ scales like the square
root of the cosmic time~$t$ and the deceleration variable~$q=-\ddot{a}/(aH^{2})\lessapprox1$
is very close to unity. Using this property the function~$f\propto H^{2}(q-1)$
would be sufficiently small to allow~$\beta/f\propto(q-1)^{-1}$
to counteract the large constant~$\Lambda$ in a similar way as at
late times. Also here $\rho_{c},\rho_{m}\ll|\Lambda|$ can be neglected
in the Friedmann equation yielding $\beta/(H^{2}(q-1))+\Lambda=0$.
As a result, the large value of~$H$ in the radiation era implies~$q\approx1$
as the solution of this equation, which means that the cosmological
evolution will be that of a radiation universe despite the large CC
term. Moreover, since $H$ decreases with time the deceleration~$q$
will slightly change its value but it stays very close to unity. It
is clear that replacing $(q-1)$ in~$f$ by $(q-\frac{1}{2})$ yields
a cosmological expansion behaviour with~$q\approx\frac{1}{2}$ for
large Hubble rates. Consequently, the effect of the CC is neutralised
also in the matter era, where~$q=\frac{1}{2}$ follows from the scale
factor $a\propto t^{2/3}$. 

In summary, a large CC can be relaxed in all major epochs of a Big
Bang universe by constructing an adequate function~$f$ in Eq.~(\ref{eq:LXCDM-Friedmann}).
In Ref.~\cite{Bauer:2009ke} the following model was proposed,\begin{equation}
f=\frac{4(2-q)}{(1-q)}\left(\frac{1}{2}-q\right)H^{2}+y\,72(1+q^{2})(1-q)H^{6},\label{eq:LXCDM-f-full}\end{equation}
which allows the relaxation of the large CC in all cosmological epochs.
The different powers of~$H$ in~$f$ ensures the correct temporal
sequence. For very large~$H$ the last term $\propto H^{6}(1-q)$
is responsible for the radiation era, while the first term~$\propto H^{2}(\frac{1}{2}-q)$
relaxes the CC in the matter and the final de~Sitter eras. The parameter~$y$
is related to the radiation-matter transition. Note that Eq.~(\ref{eq:LXCDM-f-full})
can be written in terms of the Ricci scalar~$R=6H^{2}(1-q)$, the
squared Ricci tensor~$Q=R_{ab}R^{ab}=12H^{4}(q^{2}-q+1)$ and the
squared Riemann tensor~$T=R_{abcd}R^{abcd}=12H^{4}(q^{2}+1)$, i.e.\
$f=(R^{2}-Q+yR^{2}T)/R$. Moreover, the model features an implicit
interaction with dark matter~\cite{Grande:2006nn} as well as interesting
tracking properties~\cite{Bauer:2009ke} and a reasonable evolution
of perturbations~\cite{Bauer:2009jk}.

\section{The relaxation mechanism in modified gravity}

\label{sec:Metric}In the previous section the large term~$\Lambda$
was enhanced directly by a dynamical term~$\beta/f$, which provides
a reasonable cosmological evolution. Now, we show that this relaxation
mechanism can be implemented in an action functional for modified
gravity. Consider the action\begin{equation}
\mathcal{S}=\int d^{4}x\sqrt{|g|}\left[\frac{R}{16\pi G_{N}}+\mathcal{L}_{\text{mat}}-\Lambda-\beta\, F(R,G)\right],\label{eq:Metric-Action}\end{equation}
which contains the Einstein-Hilbert term, the matter Lagrangian~$\mathcal{L}_{\text{mat}}$,
the large CC term~$\Lambda$ and the modified gravity functional~$F(R,G)$
in terms of the Ricci scalar and the Gau\ss-Bonnet invariant~$G=R^{2}-4Q+T$.
As before $\beta$ is a constant parameter. In the metric formalism
the variation of~$\mathcal{S}$ with respect to the metric~$g_{ab}$
yields the Einstein equations\begin{equation}
G_{ab}=-8\pi G_{N}\left[T_{ab}+g_{ab}\Lambda+2\beta E_{ab}\right],\label{eq:Metric-EinsteinEq}\end{equation}
where $T_{ab}$ is the energy-stress tensor of matter, and~$E_{ab}$
emerges from the new term~$F(R,G)$. On a flat Friedmann-Lema\^itre-Robertson-Walker
(FLRW) background~$E_{ab}$ is completely defined by the effective
energy density~$\rho_{F}$ and pressure~$p_{F}$. One finds\begin{equation}
\rho_{F}=2\beta\left[\frac{1}{2}F-3(\dot{H}+H^{2})F^{R}+3H\dot{F}^{R}-12H^{2}(\dot{H}+H^{2})F^{G}+12H^{3}\dot{F}^{G}\right],\label{eq:Metric-rhoF}\end{equation}
where $F^{R,G}$ correspond to partial derivatives of~$F$ with respect
to~$R$ and~$G$, respectively. Next, we consider the model from
Ref.~\cite{Bauer:2009ea}, where $F(R,G)=1/B$ with~$B:=\frac{2}{3}R^{2}+\frac{1}{2}G+(y\, R)^{3}$
is very similar to~$f$ in Eq.~(\ref{eq:LXCDM-f-full}) if written
in terms of~$H$ and~$q$,\begin{equation}
B=24H^{4}\left(q-\frac{1}{2}\right)(q-2)+H^{6}\left[6y(1-q)\right]^{3}.\label{eq:Metric-B}\end{equation}
Since every derivative of~$F$ in Eq.~(\ref{eq:Metric-rhoF}) yields
another factor~$B$ in the denominator of~$\rho_{F}$ we obtain
$\rho_{F}=N/B^{d}$ with $d\ge1$. $N$ denotes terms involving~$H$
and its time derivatives. As a result, the structure of the energy
density~$\rho_{F}$ is very similar to~$\beta/f$ from the previous
section and it behaves in a similar way. $\rho_{F}$ becomes sufficiently
large in the radiation, matter and respectively the late-time DE eras
because the terms $(q-1)$, $(q-\frac{1}{2})$ and $H^{4}$ in the
denominator~$B$ are small in the corresponding epochs. Therefore,
the Friedmann equation reduces to~$\rho_{F}+\Lambda=0$ with corrections
much smaller than~$|\Lambda|$. As before, this yields in a dynamical
way a relaxed cosmological expansion behaviour despite the large term~$\Lambda$.
Note that the numerator~$N$ in~$\rho_{F}$ does not harm the working
principle of the CC relaxation mechanism, but it provides new solutions
for the late-time behaviour, e.g.\ accelerating power-law expansion
and future singularities. We refer the reader to Ref.~\cite{Bauer:2010wj}
for a detailed study of the above model and its generalisations.

\section{Filtering out the CC in the Palatini formalism}

\label{sec:Palatini}In the previous section we discussed the CC relaxation
mechanism via modified gravity in the metric formalism, where the
connection~$\Gamma_{bc}^{a}$ is the Levi-Civita connection of the
metric~$g_{ab}$. Now, we modify gravity in the Palatini formalism
in which~$g_{ab}$ and the connection~$\Gamma$ are treated independently.
Recently, we have shown that this property allows the construction
of a filter for the CC~\cite{Bauer:2010bu}, which is based on the
action\begin{equation}
\mathcal{S}=\int d^{4}x\,\sqrt{|g|}\left[\frac{1}{2}f(R,Q)+\mathcal{L}_{\text{mat}}[g_{ab}]-\Lambda\right],\label{eq:Palatini-Action}\end{equation}
where~$\mathcal{L}_{\text{mat}}$ does not depend on $\Gamma$. The
Ricci scalar $R=g^{ab}R_{ab}$ and~$Q=R^{ab}R_{ab}$ are quantities
which depend on~$g_{ab}$ and~$\Gamma$, while the Ricci tensor~$R_{ab}=\Gamma_{ab,e}^{e}-\Gamma_{eb,a}^{e}+\Gamma_{ab}^{e}\Gamma_{fe}^{f}-\Gamma_{af}^{e}\Gamma_{eb}^{f}$
is defined only by~$\Gamma$. Here, we assume that~$\Gamma$ and
$R_{ab}$ are symmetric, which is not the most general case~\cite{Vitagliano:2010pqsr}.
The variation of~$\mathcal{S}$ with respect to~$g_{ab}$ yields
the Einstein equations\begin{equation}
f^{R}R_{mn}+2f^{Q}R_{m}^{\,\,\, a}R_{an}-\frac{1}{2}g_{mn}f=T_{mn},\label{eq:Palatini-EOM1}\end{equation}
where we include the large CC~$\Lambda$ in the stress tensor~$T_{mn}$
in addition to ordinary matter with the energy density~$\rho$ and
the pressure~$p$. In the Palatini formalism the variation $\delta\mathcal{S}/\delta\Gamma_{bc}^{a}=0$
provides the equation for the connection\begin{equation}
\nabla_{a}[\Gamma]\left[\sqrt{|g|}(f^{R}g^{mn}+2f^{Q}R^{mn})\right]=0,\label{eq:Palatini-EOM2}\end{equation}
where $\nabla_{a}$ is the covariant derivative in terms of the yet
unknown $\Gamma$. For solving this set of equations we used the method
described in Ref.~\cite{Olmo:2009xy}. In the following we summarise
the results for the CC filter model defined by the action functional\begin{equation}
f(R,Q)=\kappa R+z\,\,\,\,\text{with}\,\,\,\, z:=\beta\left(\frac{R^{\frac{2}{3}}}{B}\right)^{m},\label{eq:Palatini-fRQ}\end{equation}
where $\kappa$, $\beta$ are parameters and $m>0$ is dimensionless.
Here, the function~$B:=R^{2}-Q$ is similar to~$f$ in Sec.~\ref{sec:LXCDM},
however, note that~$R$, $Q$ and~$B$ are in general different
from their metric versions. From Eq.~(\ref{eq:Palatini-EOM1}) we
obtain two algebraic equations for the two unknowns $R$ and $B$.
First, its trace $(-2-\frac{4}{3}m)z=\kappa R-4\Lambda+3p-\rho\approx-4\Lambda$
tells us that~$z$ is approximately constant and of the order of
the large CC. The second equation reads $\kappa R+r=\frac{2}{9}mz\frac{B}{R^{2}}+\epsilon$,
where we introduced~$r:=\rho+p$. $\epsilon$ denotes suppressed
corrections which are neglected in the following. Then, we combine
both equations with~$z$ from Eq.~(\ref{eq:Palatini-fRQ}) and find
\begin{equation}
\left(\kappa R+r\right)^{3}(\kappa R)^{4}=\rho_{e}^{7}:=\kappa^{4}\left(\frac{2}{9}mz\left(\frac{\beta}{z}\right)^{\frac{1}{m}}\right)^{3},\label{eq:kR-PolyEq}\end{equation}
indicating that the Ricci scalar~$R$ can be expressed only by the
energy density~$\rho$ and the pressure~$p$ of matter in $r$ and
by the constant~$\rho_{e}$. Moreover, we observe that vacuum energy
terms with equation of state~$p=-\rho$ do not contribute to~$R$.
Obviously, the large CC~$\Lambda$ is filtered out, it appears only
in the constant~$\rho_{e}$, which can be adjusted by the parameter~$\beta$.

In the following we restrict the cosmological discussion to negative
values of~$\rho_{e}$. In the early universe, when $-\rho_{e}\ll r\ll|\Lambda|$
we find $\kappa R=-r+\mathcal{O}(\epsilon)$ from Eq.~(\ref{eq:kR-PolyEq}).
This implies that the parameter~$\kappa$ is negative since $r>0$
for ordinary matter. At late times, when $r\ll-\rho_{e}$, the Ricci
scalar becomes constant, $\kappa R=\rho_{e}-\frac{3}{7}r+\mathcal{O}(\epsilon)$.
Next, let us express the matter energy density by a power-law in the
cosmic scale factor, $\rho\propto a^{-3(\omega+1)}$, with the matter
equation of state~$\omega$. Thus,~$R$ and~$B$ can be expressed
in terms of~$a(t)$, which allows calculating the Palatini connection~$\Gamma$
via Eq.~(\ref{eq:Palatini-EOM2}) as a function of~$a(t)$ and its
derivatives. From~$\Gamma$ the Ricci tensor~$R_{ab}[\Gamma]$ can
be determined and the corresponding Ricci scalar~$R=g^{ab}R_{ab}[\Gamma]$
must be equal to~$R(r)$, which we found from Eq.~(\ref{eq:kR-PolyEq}).
With the ansatz $a(t)\propto t^{s}$ for the scale factor we obtain
the following equations for the Hubble rate~$H.$ In the era where~$r$
is dominated by dust matter with~$\omega=0$ we find $(-\kappa)H^{2}=\frac{4}{27}\rho_{\text{dust}}$,
whereas in the radiation dominated epoch with $\omega=\frac{1}{3}$
we obtain $(-\kappa)H^{2}=\frac{4}{25}\rho_{\text{radiation}}$. Obviously,
these results are very close to general relativity when the parameter~$\kappa\sim1/G_{N}$
is suitably chosen. Moreover, at late times we find the de~Sitter
solution $\kappa R=\kappa(3H^{2})=\rho_{e}$, where the late-time
Hubble rate can be adjusted by the parameter~$\beta$ in~(\ref{eq:kR-PolyEq}).
As in the previous sections it is not necessary to fine-tune its value.
Note in the solutions above the cosmic expansion is controlled by
matter only, but not by the large CC term, which has been filtered
out. It can be shown that the CC filter is active in black hole-like
environments, too~\cite{Bauer:2010bu}. As a closing remark, our
setup in the Palatini formalism leads to second order equations of
motion just as general relativity.

\section{Conclusions}

We have discussed several arguments for the existence of a presumed
large CC, which has to be neutralised to permit a reasonable evolution
of the universe. We have also argued that naively replacing the cosmological
term by other sources for late-time acceleration corresponds to the
introduction of a fine-tuned counterterm. As an alternative to this
method we have reviewed three recent approaches to relax the cosmic
expansion in the presence of a large CC.

\section*{Acknowledgments}

This work has been supported by DIUE/CUR Generalitat de Catalunya
under project 2009SGR502, by MEC and FEDER under project FPA2007-66665
and by the Consolider-Ingenio 2010 program CPAN CSD2007-00042.

\section*{References}

\bibliographystyle{iopart-num-href}
\bibliography{Palatini-Relax-Pascos}

\providecommand{\newblock}{}
\begin{thebibliography}{10}
\expandafter\ifx\csname url\endcsname\relax
  \def\url#1{{\tt #1}}\fi
\expandafter\ifx\csname urlprefix\endcsname\relax\def\urlprefix{URL }\fi
\providecommand{\eprint}[2][]{\href{http://arxiv.org/abs/#2}{{\ttfamily #2}}}

\bibitem{Weinberg:1988cp}
Weinberg S 1989 {\em Rev. Mod. Phys.\/} {\bf 61} 1--23
  \href{http://dx.doi.org/10.1103/RevModPhys.61.1}{(\textit{Link to journal})}

\bibitem{Bauer:2010wj}
Bauer F, Sola J and Stefancic H 2010  (\textit{Preprint} \eprint{1006.3944})

\bibitem{Copeland:2006wr}
Copeland E~J, Sami M and Tsujikawa S 2006 {\em Int. J. Mod. Phys.\/} {\bf D15}
  1753--1936 (\textit{Preprint} \eprint{hep-th/0603057})

\bibitem{Amanullah:2010vv}
Amanullah R {\em et~al.\/} 2010 {\em Astrophys. J.\/} {\bf 716} 712--738
  (\textit{Preprint} \eprint{1004.1711})

\bibitem{Nojiri:2006ri}
Nojiri S and Odintsov S~D 2006 {\em ECONF\/} {\bf C0602061} 06
  (\textit{Preprint} \eprint{hep-th/0601213})

\bibitem{Stefancic:2008zz}
Stefancic H 2009 {\em Phys. Lett.\/} {\bf B670} 246--253 (\textit{Preprint}
  \eprint{0807.3692})

\bibitem{Bauer:2009ke}
Bauer F, Sola J and Stefancic H 2009 {\em Phys. Lett.\/} {\bf B678} 427--433
  (\textit{Preprint} \eprint{0902.2215})

\bibitem{Grande:2006nn}
Grande J, Sola J and Stefancic H 2006 {\em JCAP\/} {\bf 0608} 011
  (\textit{Preprint} \eprint{gr-qc/0604057})

\bibitem{Bauer:2009jk}
Bauer F 2010 {\em Class. Quant. Grav.\/} {\bf 27} 055001 (\textit{Preprint}
  \eprint{0909.2237})

\bibitem{Bauer:2009ea}
Bauer F, Sola J and Stefancic H 2010 {\em Phys. Lett.\/} {\bf B688} 269--272
  (\textit{Preprint} \eprint{0912.0677})

\bibitem{Bauer:2010bu}
Bauer F 2010  (\textit{Preprint} \eprint{1007.2546})

\bibitem{Vitagliano:2010pqsr}
Vitagliano V, Sotiriou T~P and Liberati S 2010  (\textit{Preprints}
  \eprint{1007.3937}, \eprint{1008.0171})

\bibitem{Olmo:2009xy}
Olmo G~J, Sanchis-Alepuz H and Tripathi S 2009 {\em Phys. Rev.\/} {\bf D80}
  024013 (\textit{Preprint} \eprint{0907.2787})

\end{thebibliography}

\end{document}